\documentclass[a4paper,10pt]{article}
\usepackage{epsf,latexsym,amsmath}
\usepackage{times}
\usepackage{amssymb}
\usepackage{graphicx}

 \graphicspath{/img}
 \DeclareGraphicsExtensions{.pdf}
 \pdfoutput=1

 \usepackage{subfigure}

\newcommand*{\diff}{\mathop{}\!\mathrm{d}}
\newcommand{\iseij}[6]{\left\{
    \begin{array}{ccc}
      #1&#2&#3\\
      #4&#5&#6
    \end{array}\right\}}

\begin{document}

\begin{center}
{ \Large \bf  Exact and asymptotic computations of elementary spin networks:
        classification of the quantum--classical boundaries $\,(\ddag)$}
\end{center}

\vspace{24pt}

\begin{center}
{ \large{\sl A.C.P. Bitencourt$^{\,(1)}$,
A. Marzuoli$^{\,(2)}$,
M. Ragni$^{\,(3)}$,
R.W. Anderson$^{\,(4)}$ and V. Aquilanti$^{\,(5)}$}}\\
\end{center}

\vspace{1cm}

{\small
\noindent (1) Centro de Ci$\hat{\textnormal{e}}$ncias Exatas e  Tecnol\'ogicas,
Universidade Federal do Rec$\hat{\textnormal{o}}$ncavo da Bahia, 
Cruz das Almas, Bahia (BR); 
(2) Dipartimento di Matematica `F Casorati',
Universit\`a degli Studi  di Pavia and INFN, Sezione di Pavia, 27100 Pavia (IT);
(3) Departamento de F\'isica,
Universidade Estadual de Feira de Santana, Feira de Santana, Bahia (BR);
(4) Department of Chemistry,
University of California, Santa Cruz, California (USA);
\noindent (5) Dipartimento di Chimica,
Universit\`a degli Studi di Perugia, 06123 Perugia (IT)}
                         
 \vspace{24pt}

\noindent {\bf Abstract}\\

Increasing interest is being dedicated in the last few years to
the issues of exact computations and  asymptotics of spin
networks. The large--entries regimes (semiclassical limits) occur
in many areas of physics and chemistry, and in particular in
discretization algorithms of applied quantum mechanics. Here we
extend recent work on the basic  building block of spin networks,
namely the Wigner $6j$ symbol or Racah coefficient, enlightening
the insight gained by exploiting its self--dual properties and
studying it as a function of two (discrete) variables. This arises
from its original definition as an (orthogonal) angular momentum
recoupling matrix. Progress also derives from recognizing its role
in the foundation of the modern theory of classical orthogonal
polynomials, as extended to include discrete variables. Features
of the imaging of various regimes of these orthonormal matrices
are made explicit by computational advances --based on traditional
and new recurrence relations-- which allow an interpretation of
the observed behaviors in terms of an underlying Hamiltonian
formulation  as well. This paper provides a contribution to the
understanding of the transition between two extreme modes of the
$6j$, corresponding to the nearly classical and the fully quantum
regimes, by studying the boundary lines (caustics) in the plane of
the two matrix labels. This analysis marks the evolution of the
turning points of relevance for the semiclassical regimes and puts
on stage an unexpected key role of the Regge symmetries of the
$6j$.

\vspace{44pt}

{\small $(\ddag)$
Talk presented at ICCSA 2012 [12th International Conference on Computational Science 
and Applications, Salvador de Bahia (Brazil) June 18-21, 2012],\\
Lecture Notes in Computer Science {\bf 7333}, Part I, pp 723-737 (Murgante B. et al. Eds.)
Springer-Verlag,  Berlin-Heidelberg 2012 ISBN 978-3-642-31124-6}

 \vfill
\newpage                        




\section{Introduction}
This is an account of progress on understanding elementary spin
networks in view of their ubiquitous occurrence in applications
far beyond the theory of angular momentum in quantum mechanics
where they were introduced by Wigner, Racah and others. They have
been (and are) precious tools for computational nuclear, atomic,
molecular and chemical physics: best known  examples  are vector
coupling and recoupling coefficients and $3nj$ symbols
 \cite{ACG.96, AC.00, ACC.01, AC.01, FCV.03, AHLY.07,ABFMR.08,AAF.08, ABFMR.09, AAM.09,
 RBFAAL.10,AHHJLY.11,AQU95:15694,AQU98:3792,AQU01:103}.

The diagrammatic tools for ``spin networks'' were developed by the
Yutsis school and by others \cite{YuLeVa,Russi}, and were given
this collective name in connection with applications to
discretized models for quantum gravity after Penrose \cite{Pen},
Ponzano and Regge \cite{PonzRegge}, and many others (see {\em
e.g.} \cite{Rov,CaMaRa}).

The basic building blocks of (all) spin networks are the Wigner
$6j$ symbols or Racah coefficients, which we are study here by
exploiting their self dual properties and consequently looking at
them as  functions of two variables. This approach is most natural
in view of their origin as matrix elements describing recoupling
between alternative angular momentum binary coupling schemes, or
between alternative hyperspherical harmonics, or between
alternative atomic and molecular orbitals, {\em etc.}, and
utilizes progress in understanding their role in the foundation of
the modern theory of classical orthogonal polynomials, as extended
to include discrete variables \cite{Askey}. Features of the
imaging of the orthonormal matrices is made possible by
computational advances, that permit to elaborate accurate
illustrations and comparisons, using exact computations based on
traditional and new recurrence relations, which allow in turn an
interpretation of the observed behaviors in terms of an underlying
Hamiltonian formulation. An unexpected key role of the
(mysterious) Regge symmetries \cite{ReSymm} of the $6j$ is briefly
discussed. Suitable use of the results for discretization
algorithms of applied quantum mechanics is stressed, with
particular attention to problems arising in atomic and molecular
physics.

For background information and notation, we will refer to our
previous papers \cite{ABFMR.08, ABFMR.09, RBFAAL.10}, regarding
the one dimensional view, and also to \cite{RobYu,JPA2012}, where
ample attention is devoted to the two dimensional perspective and
where the related $4j$ model is elaborated in great  detail.

\section{Background information}

Semiclassical and asymptotic views are introduced to describe the
dependence on parameters. They originated from the early
association due to Racah and Wigner  to geometrical features,
respectively a dihedral angle and the volume of an associated
tetrahedron (see Fig. 4 in \cite{RBFAAL.10}), which is the
starting point of the seminal paper by Ponzano and Regge
\cite{PonzRegge}. Their results provided an impressive insight
into the functional dependence of angular momentum functions
showing a quantum mechanical picture in terms of formulas which
describe classical and non--classical discrete wavelike regimes,
as well as the transition between them.

The symmetries with respect to exchange of role of the matrix
entries of the $6j$ (transpositions), easily understood also from
the associated tetrahedron picture, are well characterized.
Similarly, we find that the mirror symmetries can be exploited
when the symbols are employed in applications beyond angular
momentum theory, implying meaning of the entries as quantities
that can have a sign.

From the detailed study of combinatorial properties (requiring
only triangular relationships associated to angular momentum
coupling, and the closure quadrilateral property associated to
coupling of four angular momenta), it emerged very recently
\cite{RobYu} how to uniquely assign labels to the elements of the
grids for the matrix for which the orthonormalized Racah--Wigner
coefficients are the elements. Figures of zero volume and ridges
are among properties that can be monitored on such square
"screen": we sketch relationships and concept of Regge twins,
``canonical'' form and simplification of the following due to a
suitable chosen ordering. From a computational viewpoint, explicit
formulas are available as sums over a single variable. However,
resorting to recursion formulas appears most convenient for exact
calculations. Also, we will exploit them for semi--classical
analysis, both to understand the high $j$--limit and, in reverse,
to interpret the symbols them as discrete wave--functions obeying
Schr\"odinger type of difference (rather then differential)
equations. We have derived and computationally implemented a
two--variable recurrence that permits construction of the whole
orthonormal matrix. The derivation follows our paper in
\cite{AAM.09} and is of interest also for other $3nj$ symbols in
general. Separation of the two--variable recurrence relation leads
to the basic three--term recurrence as depending on a single
variable. It can be shown that it can be re-derived also from the
Biedenharn--Elliot relationship in a form that shows the
connection to a Schr\"odinger type of equation in the Hamiltonian
formulation (for alternative Lagrangian formulations, see
\cite{JPA2012}). The following analysis of the caustics (and of
the ridge curves, see below) is intended to characterize the modes
of the spin network, as well as the guiding principles of the
asymptotic analysis.

\section{\label{sec.02new}The screen: mirror, Piero and Regge symmetries}

Following e.g. the Schulten and Gordon approach \cite{SchGorA}, in
\cite{SchGorB} it is shown that the $6j$ symbol becomes the
eigenfunction of the Schr\"odinger--like equation in the variable
$q$, a continuous generalization of  $j_{12}$:
 \begin{equation}
  \left[\frac{\diff}{\diff q^2} +p^2(q)\right]\Psi(q)=0~,
 \end{equation}
where $\Psi(q)$ is related to
 \begin{equation}
  \iseij{j_1}{j_2}{j_{12}}{j_3}{j}{j_{23}}~
 \end{equation}
and $p^2$ is related with the square of the volume $V$ of the
associated tetrahedron (Fig 4 in \cite{RBFAAL.10}), whose edges
are considered continuous and given by
 $J_1=j_1+1/2$, $J_2=j_2+1/2$, $J_{12}=j_{12}+1/2$, $J_3=j_3+1/2$, $J=j+1/2$,
and $J_{23}=j_{23}+1/2$. The Cayley--Menger determinant permits to
calculate the square of the volume of a generic tetrahedron in
terms of (squares of) its edge lengths according to
 \begin{equation}\label{Vol}
  V^2=\frac{1}{288}\left|\begin{array}{ccccc}
      0       & J^2_3    & J^2      & J^2_{23} & 1\\
      J^2_3   & 0        & J^2_{12} & J^2_2    & 1\\
      J^2     & J^2_{12} & 0        & J^2_1    & 1\\
      J^2_{23}& J^2_2    & J^2_1    & 0        & 1\\
      1       & 1        & 1        & 1        & 0
    \end{array}\right|~.
 \end{equation}
The condition for the tetrahedron with fixed edge lengths to exist
as a polyhedron in  Euclidean 3-space amounts to require $V^2 >0$,
while the $V^2=0$ and $V^2<0$ cases were associated by Ponzano and
Regge to ``flat'' and nonclassical tetrahedral configurations
respectively. Equivalent to (\ref{Vol}) is the Gramian
determinant, used in \cite{RobYu} and \cite{JPA2012}, which
embodies a clearer relationship with a vectorial picture.

Major insight is provided by plotting both $6j$'s and geometrical
functions (volumes, products of face areas, {\em etc.}) of the
associated tetrahedra in a 2-dimensional $j_{12}-j_{23}$ plane
(the square ``screen'' of allowed ranges of $j_{12}$ and $j_{23}$
to be used in all the pictures below). Actually both (\ref{Vol})
and the Gramian are equivalent to the famous formula known to
Euler but first found five centuries ago by the Renaissance
mathematician, architect and painter Piero della Francesca. The
formulas of course embody all the well known ``classical''
symmetries of geometrical tetrahedra, which  show up in the $6j$
symbol as well on applying concerted interchanges of its entries.
Non-obvious symmetries of particular relevance in what follows are
listed below.

\begin{itemize}
\item[(i)] {\em The mirror symmetry}. The appearance of squares of
tetrahedron edges entails that the invariance with respect to the
exchange $J \leftrightarrow - J$ implies formally $j
\leftrightarrow -j-1$ with respect to the entries of the $6j$
symbol. Although this is physically irrelevant when the $j$'s are
pseudo--vectors, such as physical spins or orbital angular
momenta, it can be of interest for other ({\em e.g.} discrete
algorithms) applications. Regarding the screen, it can be seen
that actually by continuation of the abscissa $x=J_{12}$ and
ordinate $y=J_{23}$ to negative values, one can have replicas that
can be glued by cutting out regions shaded in \cite{ABFMR.09}.
This allows mapping of the screen to the $S^2$ phase space found
in \cite{JPA2012}.

\item[(ii)] {\em Piero line.} In general, an exchange of opposite
edges of a tetrahedron (or of the two entries in a column in the
$6j$ symbol) corresponds to different tetrahedra and different
symbols. In Piero formula, there is a term due to this difference
that vanishes when any pair of opposite edges are equal. In
general, a line can be drawn on the screen when ranges of $j_{12}$
and $j_{23}$ overlap and in the  screen one may encounter what we
call a Piero line: when two entries in a column are equal, this
line is a diagonal corresponding to the exchange $x
\leftrightarrow y$. Then, images on the screen will be symmetric
with respect to such a line, as shown below with explicit
examples.

\item[(iii)] {\em Regge symmetries.} The manifestation of these
intriguing symmetries in the present context is of paramount
interest: it might be elucidated through connection with the
projective geometry of the elementary quantum of space, which is
being reconsidered from the viewpoint of association to the
polygonal inequalities (triangular and quadrilateral in the $6j$
case), which have to be enforced in any spin networks. We find
insightful to exhibit the basic Regge symmetry \cite{ReSymm} in
the following form
\end{itemize}

\begin{eqnarray}\label{eqq}
\iseij{j_1}{j_2}{j_{12}}{j_3}{j}{j_{23}} &=&
\iseij{(j_1+j_2 -j_3 +j)/2}{(j_1+j_2+j_3-j)/2}{j_{12}}{(-j_1+j_2+j_3+j)/2}
{(j_1-j_2+j_3+j)/2}{j_{23}}\nonumber\\
= & \, & \iseij{j_1 + \rho}{j_2- \rho}{j_{12}}{j_3+ \rho}{j- \rho}{j_{23}}~,
\end{eqnarray}
where $\rho= (-j_1+j_2-j_3-j)/2 = [(j_2+j)-(j_1+j_3)]/2$. Often
the Regge symmetry is written in terms of the semi--perimeter
$s=(j_1+j_2+j_3+j)/2$. Obviously,  $s= \rho +j_1 +j_3 = -\rho +j_2
+j$.

The key observation is made in \cite{JPA2012} that the range of
both $J_{12}$ and $J_{23}$,namely the ``size'' of the screen, is
given by 2min$\{J_1,J_2,J_3,J,J_1+
\rho,J_2+\rho,J_3+\rho,J+\rho\}$. Therefore, for definiteness and
no loss of generality, we are allowed in most cases  to
conveniently choose for the discussion whichever of the two $6j$
symbols twinned by (\ref{eqq}) contains such a minimum value.


\section{\label{sec.02}Features of the tetrahedron volume function}

 Looking at the volume $V$ as a function of $x=J_{12}$ and $y=J_{23}$,
 and after some algebraic manipulations,
 we get the expressions for the $x_{Vmax}$ and $y_{Vmax}$
 that correspond to the maximum of the volume for a fixed value of $x$ or $y$
(respectively), namely
 \begin{eqnarray}
  \label{eq.02.01a}x_{Vmax}^2 &=& \frac{A(J,J_2) + J^2_t\,y^2 - y^4}{2\,y^2}~,\\
  \label{eq.02.01b}y_{Vmax}^2 &=& \frac{A(J_2,J) + J^2_t\,x^2 - x^4}{2\,x^2}~,
 \end{eqnarray}
 where
 \begin{eqnarray}
  A(a,b) &=& (J_1+a)(J_1-a)(J_3+b)(J_3-b)~,\nonumber\\
  J^2_t &=& J^2_1 + J^2_2 + J^2_3 + J^2~.
 \end{eqnarray}
  We call ``ridge'' curves the  plots of Eqs. (\ref{eq.02.01a}) and (\ref{eq.02.01b}) on the $x, y$-screen.
  Each one marks configurations of the associated tetrahedron when two specific pairs of triangular 
faces are orthogonal.
  The corresponding values of the volume ($V_{max,x}$ and $V_{max,y}$) are
 \begin{eqnarray}
  V_{max,x} &=& \frac{2\,F(J_1,J,y)F(J_2,J_3,y)}{3\,y}~,\\
  V_{max,y} &=& \frac{2\,F(J_1,J_2,x)F(J,J_3,x)}{3\,x}~,
 \end{eqnarray}
 where
 \begin{equation}
  F(a,b,c)=\frac{\sqrt{(a+b+c)(a+b-c)(a-b+c)(-a+b+c)}}{4}~,
 \end{equation}
 is the area of the triangle with sides $a$, $b$ and $c$.

 Curves corresponding to  $V=0$ (the caustic curves) obey the equations
 \begin{eqnarray}
  \label{eq.02.10a}x_z^2 &=& x_{Vmax}^2 \pm \frac{12\,V_{max,x}}{y}~,\nonumber\\
  \label{eq.02.10b}y_z^2 &=& y_{Vmax}^2 \pm \frac{12\,V_{max,y}}{x}~,
 \end{eqnarray}

\begin{figure}\centering
  \subfigure[ ]{\label{fig.01.a}
          \includegraphics[width=.40\textwidth]{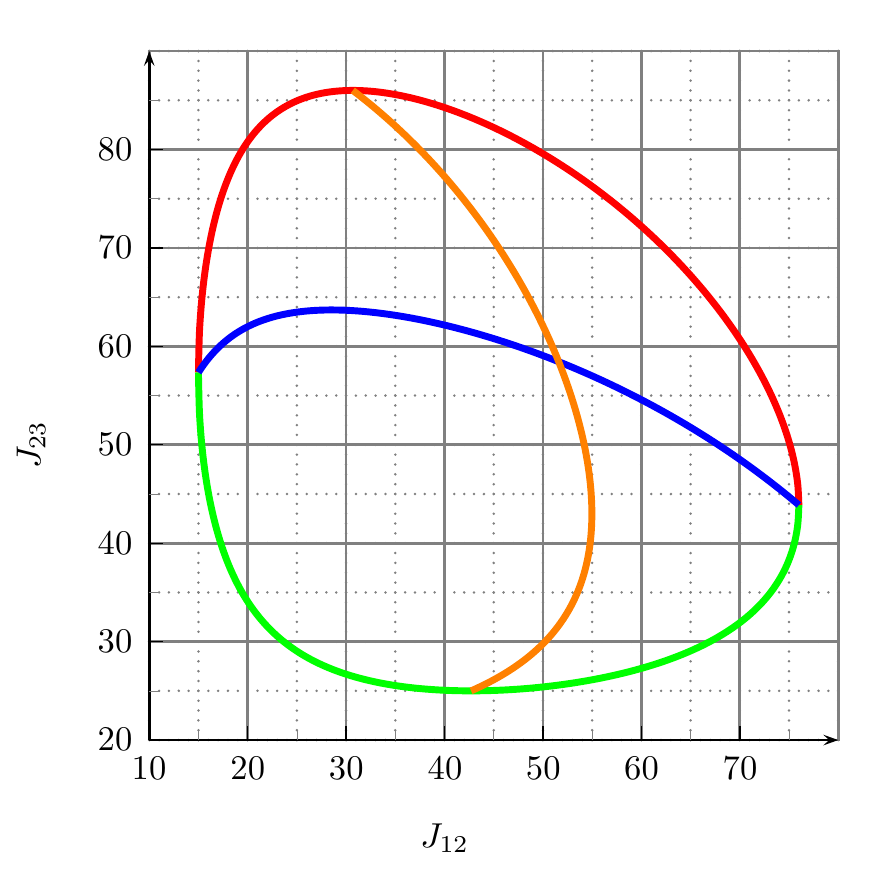}}
  \subfigure[ ]{\label{fig.01.b}
          \includegraphics[width=.40\textwidth]{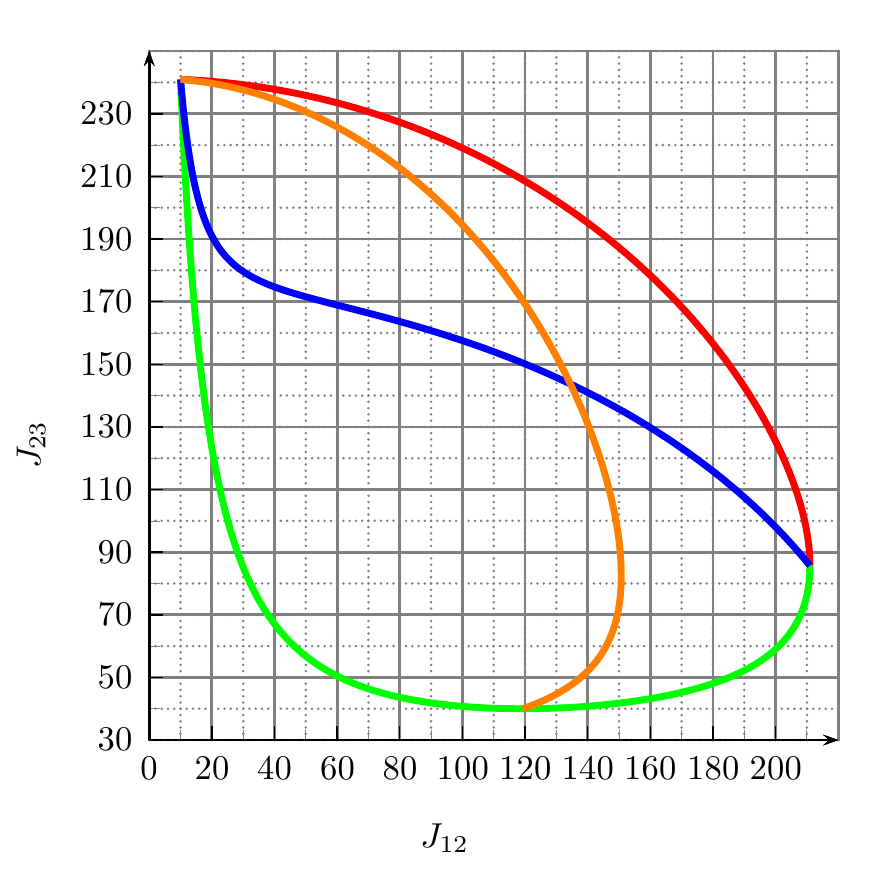}}
  \subfigure[ ]{\label{fig.01.c}
          \includegraphics[width=.40\textwidth]{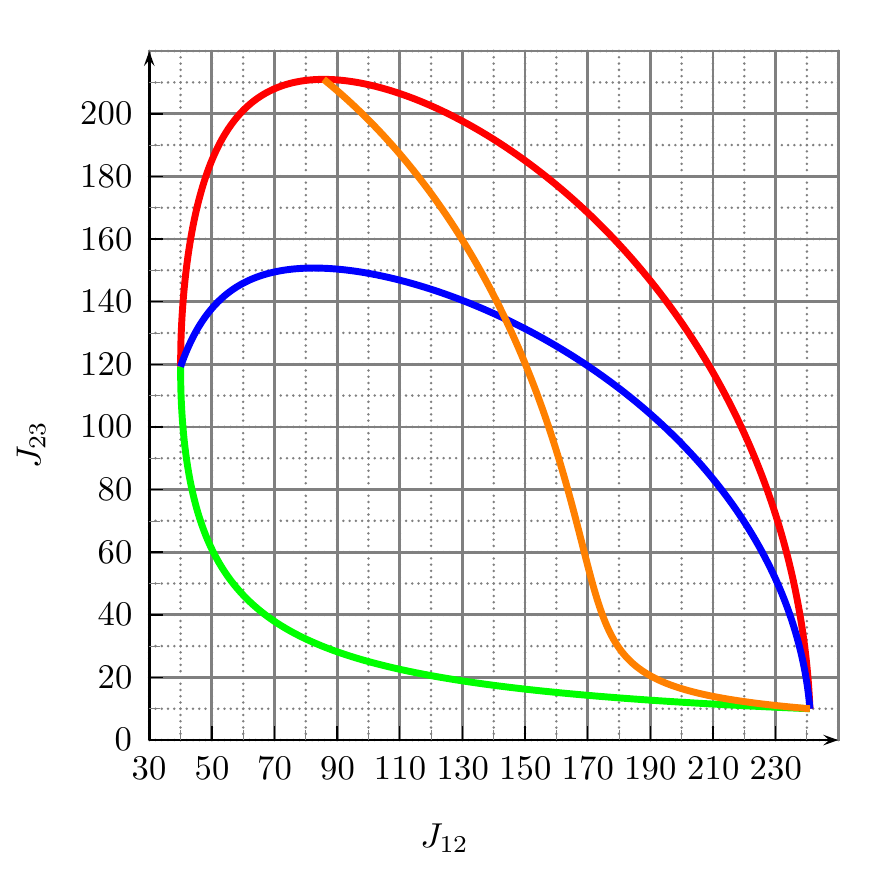}}
  \subfigure[ ]{\label{fig.01.d}
          \includegraphics[width=.40\textwidth]{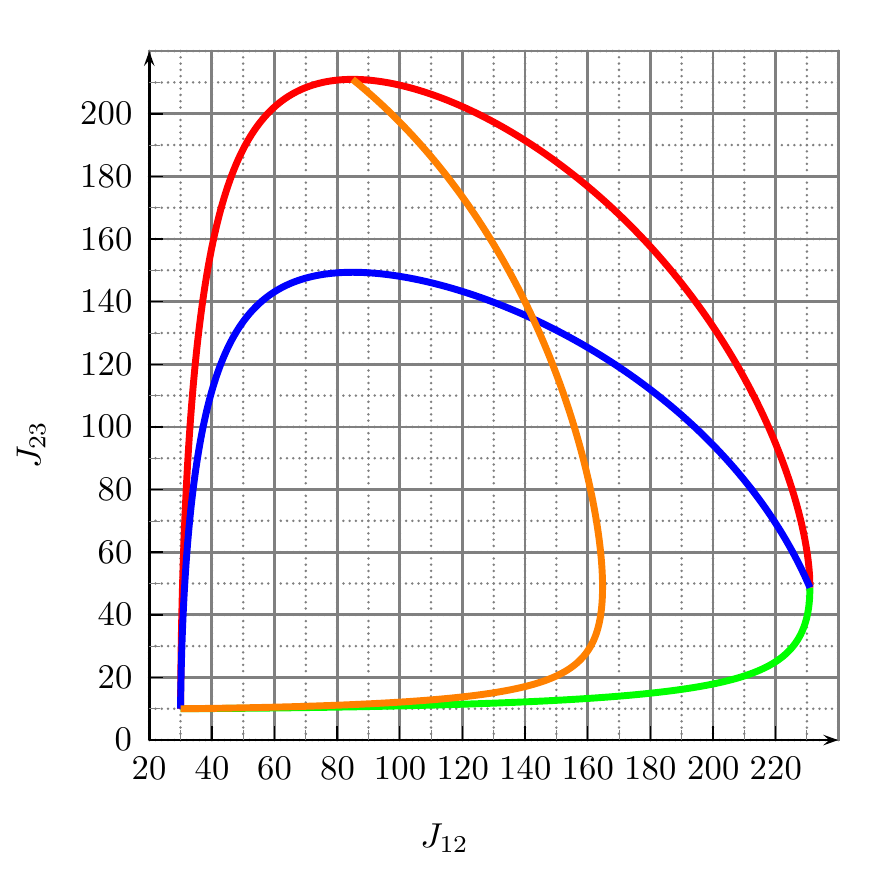}}
  \caption{\label{fig.01}Plots of the caustics and ridges given by Eqs.(\ref{eq.02.01a}-\ref{eq.02.10b})
           for four sets of $6j$-symbols.
           Panel \ref{fig.01.a}, $j_1= 45$, $j_2= 30$, $j_3= 55$ and $j= 60$. $15 \leq J_{12} \leq  76$ and $25 \leq J_{23} \leq 86$.
           Panel \ref{fig.01.b}, $j_1=140$, $j_2=130$, $j_3=110$ and $j=100$. $10 \leq J_{12} \leq 211$ and $40 \leq J_{23} \leq 241$.
           Panel \ref{fig.01.c}, $j_1=140$, $j_2=100$, $j_3=110$ and $j=130$. $40 \leq J_{12} \leq 241$ and $10 \leq J_{23} \leq 211$.
           Panel \ref{fig.01.d}, $j_1=140$, $j_2=110$, $j_3=100$ and $j=130$. $30 \leq J_{12} \leq 231$ and $10 \leq J_{23} \leq 211$.}
 \end{figure}

 Figure \ref{fig.01.a} shows these curves for the $\iseij{45}{30}{j_{12}}{55}{60}{j_{23}}$ case.
 The inside  region enclosed by the ovaloid is that of finite volume tetrahedra,
 the ovaloids themselves correspond to configurations of flattened tetrahedra, specifically
 convex planar quadrilaterals at the upper right corner,concave planar quadrilaterals
 at both the upper left and and the lower right corners, and crossed planar quadrilaterals
 at the lower left corner.

 For particular values of $j_1$, $j_2$, $j_3$ and $j$ linear configurations are
 allowed as well. These interesting degenerate cases are illustrated in the other
 panels of Fig. \ref{fig.01}. In Fig. \ref{fig.01.b} the case of $j_1+j=j_2+j_3$ is considered.
 The other interesting cases are obtained for $j_1+j_2=j_3+j$ (Fig. \ref{fig.01.c})
 and for $j_1+j_3=j_2+j$ (Fig. \ref{fig.01.d}).

 Note that, due to the symmetry properties of the $6j$ symbols, the cases $j_1+j=j_2+j_3$
 and $j_1+j_2=j_3+j$ are intrinsically equivalent. Figure \ref{fig.01.d} shows the effect of degeneracy
 with respect to the far from obvious Regge symmetry, which manifests as specularity with
 respect to the Piero line, in this case the diagonal from the origin in the square screen.

\section{\label{sec.03}Symmetric and limiting  cases}

 When some or all the $j$'s are equal, interesting features appear in the screen. Similarly
 when some are larger than others.

\subsection{\label{sec.03.a}Symmetric cases}
\begin{figure}\centering
  \includegraphics[width=1.0\textwidth]{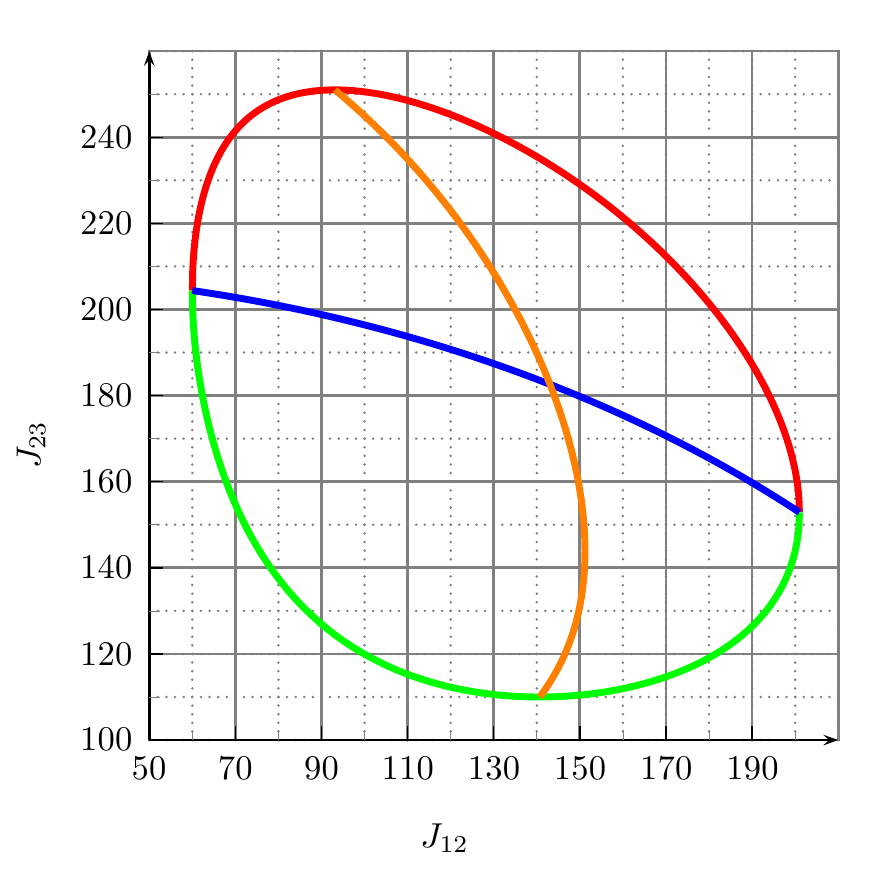}
  \caption{\label{fig.02}Plots of the caustics and ridges (Eqs.\ref{eq.02.01a}-\ref{eq.02.10b}) for the
           $6j$-symbols with $j_1= 100$, $j_2= 100$, $j_3= 150$
           and $j= 210$. $ 60 \leq J_{12} \leq 201$ and $ 110 \leq J_{23} \leq 251$.}
 \end{figure}

 For $j_1=j_2$ plots of Eqs.(\ref{eq.02.01a}-\ref{eq.02.10b}) like those of Fig. \ref{fig.02} are obtained
 (equivalent to the $j_3=j$ one, in virtue of standard symmetries).

\begin{figure}\centering
  \includegraphics[width=1.0\textwidth]{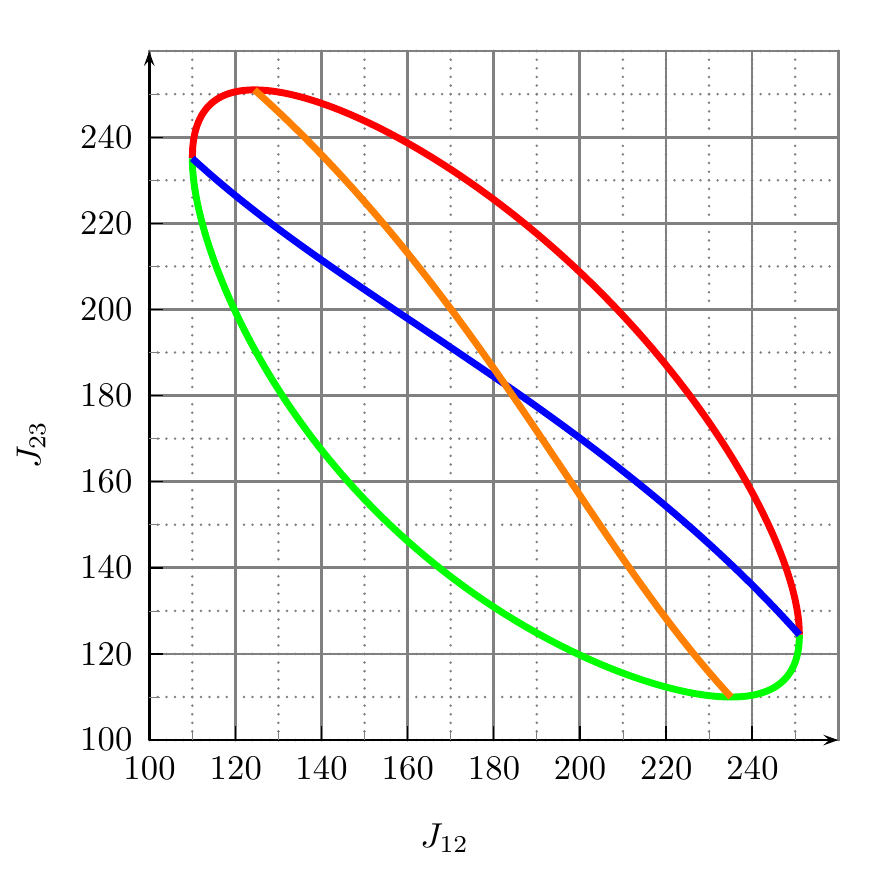}
  \caption{\label{fig.03}Plots of the caustics and ridges (Eqs.\ref{eq.02.01a}-\ref{eq.02.10b}) for the
           $6j$-symbols with $j_1= 100$, $j_2= 150$, $j_3= 100$
           and $j= 210$. $ 110 \leq J_{12} \leq 251$ and $110 \leq J_{23} \leq 251$.}
 \end{figure}

 For $j_1=j_3$ plots of Eqs.(\ref{eq.02.01a}-\ref{eq.02.10b}) like that of Fig. \ref{fig.03} are obtained.
 For symmetry this case is equivalent to the $j_2=j$ one.

\begin{figure}\centering
  \subfigure[]{\label{fig.04.a}
          \includegraphics[width=.45\textwidth]{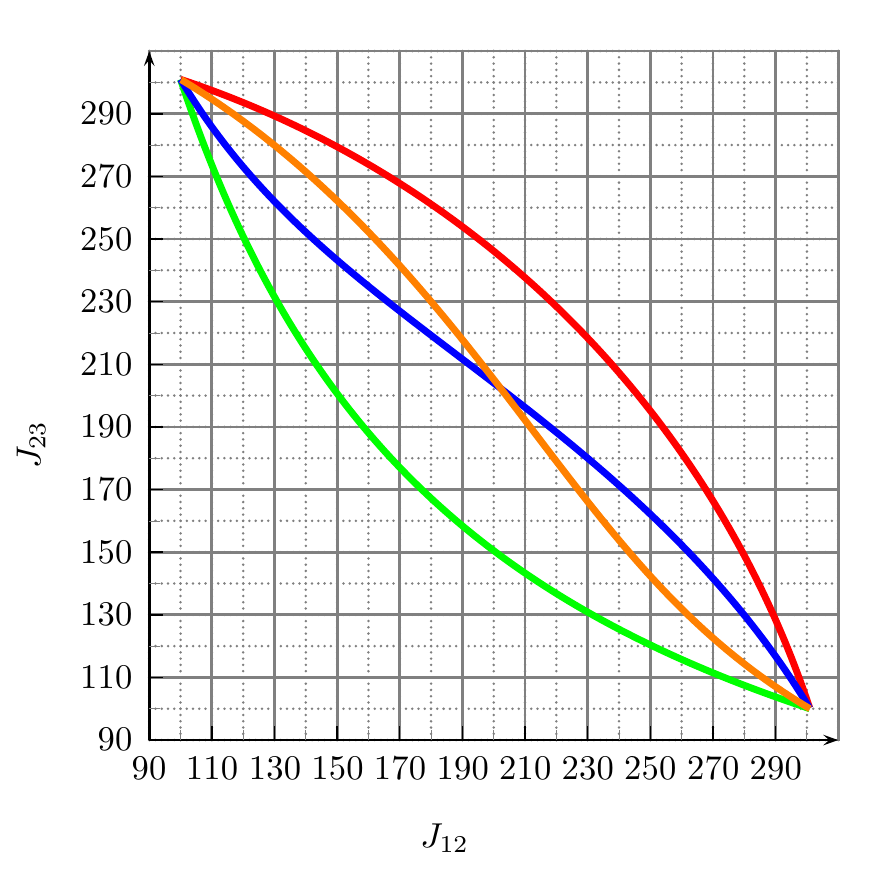}}
  \subfigure[]{\label{fig.04.b}
          \includegraphics[width=.45\textwidth]{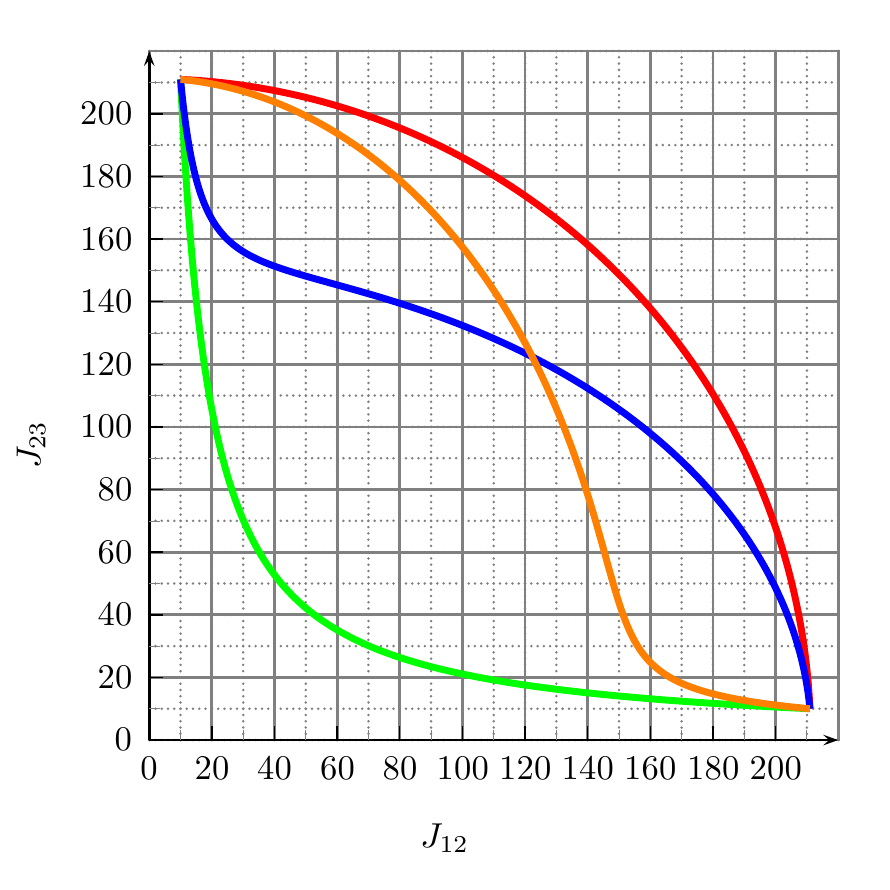}}
  \caption{\label{fig.04}Plots of the caustics and ridges (Eqs.\ref{eq.02.01a}-\ref{eq.02.10b})
           for two sets of $6j$ symbols.
           Panel \ref{fig.04.a}, $j_1= 200$, $j_2= 100$, $j_3= 200$ and $j= 100$.
           $100 \leq J_{12} \leq  301$ and $100 \leq J_{23} \leq 301$.
           Panel \ref{fig.04.b}, $j_1=110$, $j_2=100$, $j_3=110$ and $j=100$.
           $10 \leq J_{12} \leq 211$ and $10 \leq J_{23} \leq 211$.}
 \end{figure}

 Imposing both $j_1+j=j_2+j_3$ and $j_1+j_2=j_3+j$ we have that $j_1=j_3$ and $j_2=j$.
 See Figs. \ref{fig.04.a} e \ref{fig.04.b}.

\begin{figure}\centering
  \includegraphics[width=1.0\textwidth]{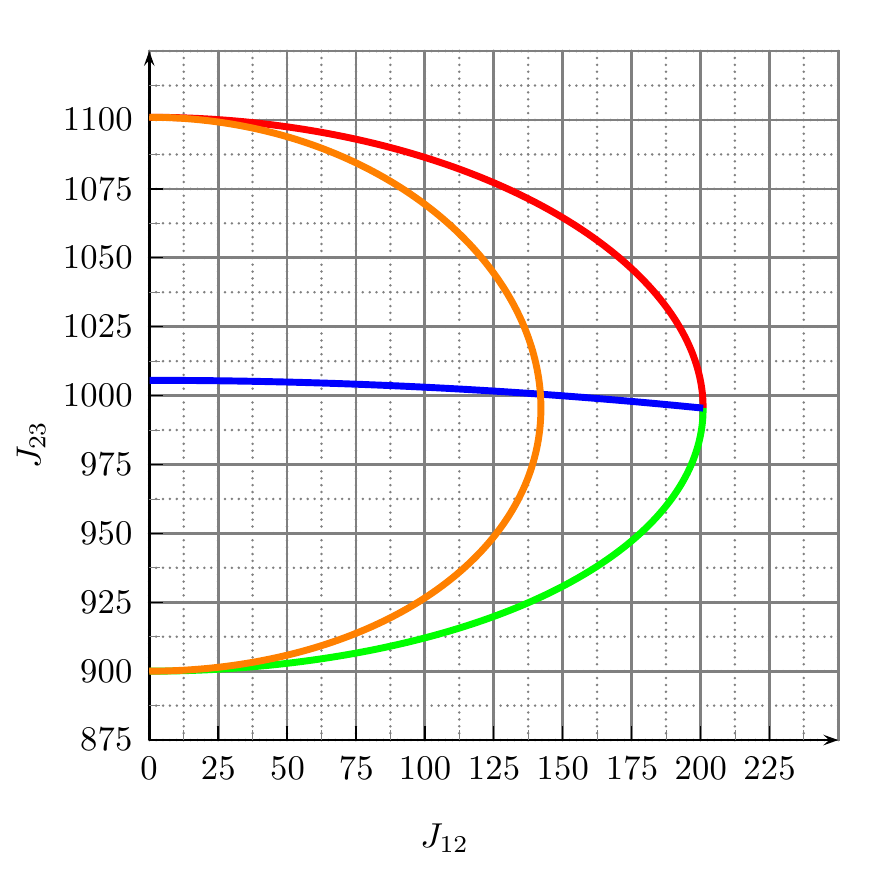}
  \caption{\label{fig.05}Plots of the caustics and ridges (Eqs.\ref{eq.02.01a}-\ref{eq.02.10b}) for the
           $6j$ symbols with $j_1= 1000$, $j_2= 1000$, $j_3= 100$
           and $j= 100$. $0 \leq J_{12} \leq 201$ and $900 \leq J_{23} \leq 1001$.
           According to the text, this figure also exemplifies features of ridges and
           caustics for a 3$j$ symbol.}
 \end{figure}

 Imposing both $j_1+j=j_2+j_3$ and $j_1+j_3=j_2+j$ we have that $j_1=j_2$ and $j_3=j$.
 In these conditions, plots of Eqs.(\ref{eq.02.01a}-\ref{eq.02.10b}) like those of Fig. \ref{fig.05}
 are obtained.

 This case is formally equivalent to the one where $j_1=j$ and $j_3=j_2$ which  is obtained
 imposing $j_1+j_2=j_3+j$ and $j_1+j_3=j_2+j$.

\subsection{\label{sec.03.b}Limiting cases}

 Interestingly, the Fig. \ref{fig.05} permits us to discuss the caustics of the $3j$ symbols as limiting case of the
 corresponding  $6j$ where three entries are larger than the other ones, namely:

 \begin{equation}
  \left(\begin{array}{ccc}
    j_1 & j_2 & j_3\\
    m_1 & m_2 & m_3
  \end{array}\right) = \lim_{R\rightarrow\infty} \iseij{j_1}{j_2}{j_3}{l_1+R}{l_2+R}{l_3+R}~,
 \end{equation}
 where
 \begin{eqnarray}
  m_1 &=& l_3-l_2~,\nonumber\\
  m_2 &=& l_1-l_3~.
 \end{eqnarray}
 If we  define
 \begin{eqnarray}
  m_1 := F+D~,\nonumber\\
  m_2 := F-D~,\nonumber\\
  \bar{R} := R+l_3-D~,
 \end{eqnarray}
 we have
 \begin{equation}
  \left(\begin{array}{ccc}
    j_1 & j_2 & j_3\\
    m_1 & m_2 & m_3
  \end{array}\right) = \lim_{\bar{R}\rightarrow\infty} \iseij{j_1}
{j_2}{j_3}{\bar{R}+F}{\bar{R}-F}{\bar{R}+D}~.
 \end{equation}
 For $\bar{R}\rightarrow\infty$ it is $\bar{R}\pm F \simeq \bar{R}$.

 The caustic of the $3j$ symbol is defined as
 \begin{equation}
  \left|\begin{array}{cccc}
      0           & J_1^2-m_1^2 & J_2^2-m_2^2 & 1\\
      J_1^2-m_1^2 & 0           & J_3^2-m_3^2 & 1\\
      J_2^2-m_2^2 & J_3^2-m_3^2 & 0           & 1\\
      1           & 1           & 1           & 0
  \end{array}\right| = 0~,
 \end{equation}
 and
 \begin{equation}
  \left|\begin{array}{cccc}
      0           & J_1^2-m_1^2 & J_2^2-m_2^2 & 1\\
      J_1^2-m_1^2 & 0           & J_3^2-m_3^2 & 1\\
      J_2^2-m_2^2 & J_3^2-m_3^2 & 0           & 1\\
      1           & 1           & 1           & 0
  \end{array}\right| = \lim_{R\rightarrow\infty}\frac{\left|\begin{array}{ccccc}
      0         & (L_1+R)^2 & (L_2+R)^2 & (L_3+R)^2 & 1\\
      (L_1+R)^2 & 0         & J^2_3     & J^2_2     & 1\\
      (L_2+R)^2 & J^2_3     & 0         & J^2_1     & 1\\
      (L_3+R)^2 & J^2_2     & J^2_1     & 0         & 1\\
      1         & 1         & 1         & 1         & 0
    \end{array}\right|}{2\,R^2}~,
 \end{equation}

 The case of caustics on the screen for four large entries, leading to reduced Wigner $d$ matrices,
 is studied in \cite{AAF.08}.

\begin{figure}\centering
  \includegraphics[width=1.0\textwidth]{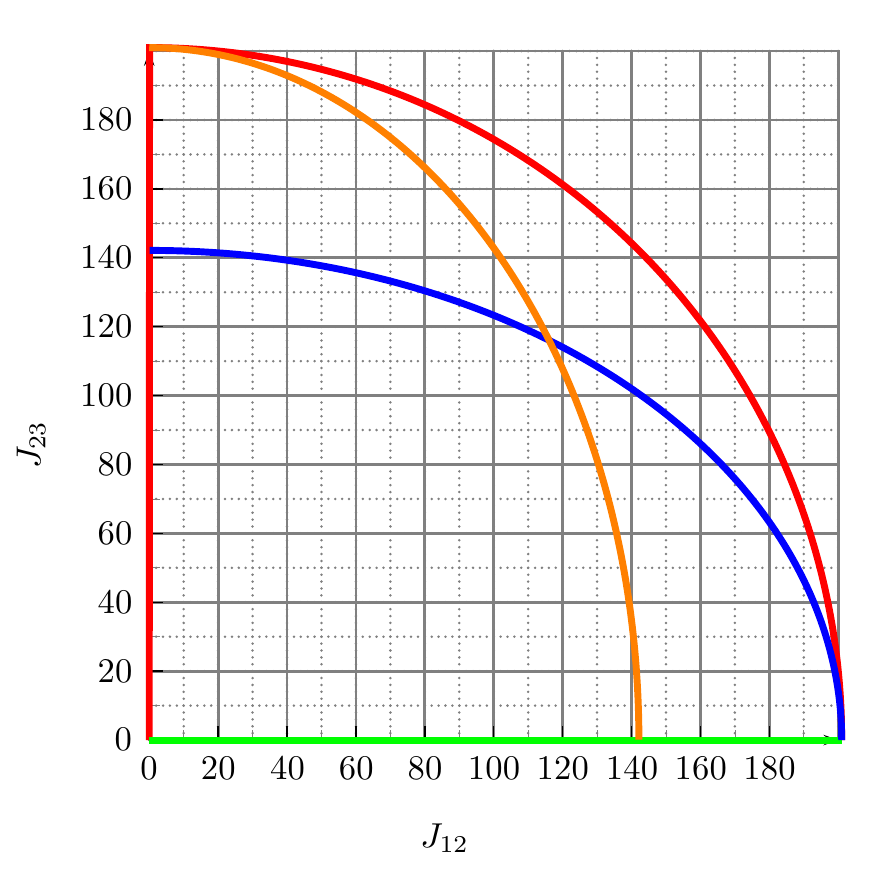}
  \caption{\label{fig.06}Plots of the caustics and ridges (Eqs.\ref{eq.02.01a}-\ref{eq.02.10b}) for the
           full symmetric $6j$-symbols with $j_1= 100$, $j_2= 100$, $j_3= 100$
           and $j= 100$. $0 \leq J_{12}, J_{23} \leq 201$. This figure clearly points out the
           interest of extending the screen by mirror symmetries.}
 \end{figure}

 For $j_1=j_2=j_3=j$ plots of Eqs.(\ref{eq.02.01a}-\ref{eq.02.10b}) like those of Fig. \ref{fig.06}
 are obtained. This case is obtained when $j_1+j=j_2+j_3$, $j_1+j_2=j_3+j$
 and $j_1+j_3=j_2+j$.

\begin{figure}\centering
  \includegraphics[width=1.0\textwidth]{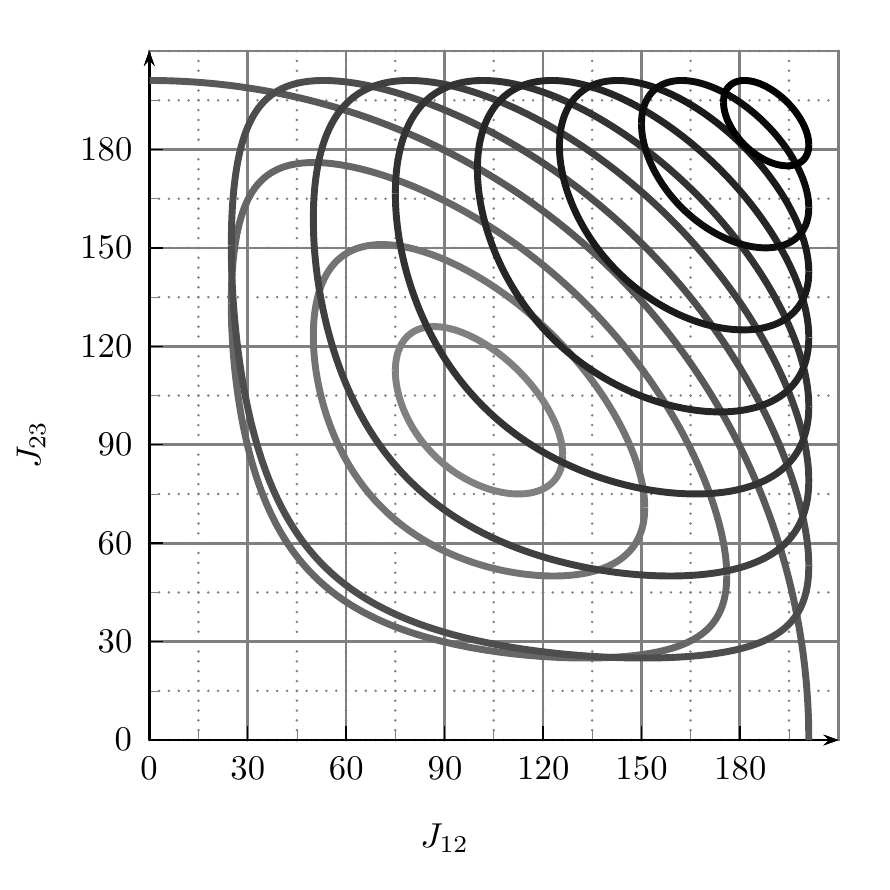}
  \caption{\label{fig.07}Plots of the caustics and ridges (Eqs.\ref{eq.02.01a}-\ref{eq.02.10b}) for
           $j_1 = j_2 = j_3 =100$ while $j$ varies from $25$ (light line) to $275$
          (dark line) with step of 25 units.}
 \end{figure}

 Fig. \ref{fig.07} shows the caustics of a full--symmetric $6j$ when the symmetry is broken
 by varying $j$.

\section{Remarks and Conclusion}
Explicit computational formulas are available either as sums over
a single variable and series. However recourse to recursion
formulas appears most convenient for both exact calculations and
-as we will emphasize- also for semiclassical analysis, in order
to understand high $j$ limits and in reverse to interpret them as
discrete wavefunctions obeying Schr\"{o}dinger type of difference
(rather then differential) equations. In further work, we derive
and computationally implement the two variable recurrence that
permits construction of the whole orthonormal matrix The
derivation follows our paper in \cite{AAM.09} and is of interest
also for other 3nj symbols in general.

The extensive images of the exactly calculated 6$j$'s on the
square screens illustrate how the caustic curves studied in this
paper separate the classical and nonclassical regions, where they
show wavelike and evanescent behaviour respectively. Limiting
cases, and in particular those referring to  3$j$ and Wigner's $d$
matrix elements can be analogously depicted and discussed.
Interesting also are the ridge lines, which separate the images in
the screen tending to qualitatively different flattening  of the
quadrilateral, namely convex in the upper right region,  concave
in the upper left and lower right ones, and crossed in the lower
left region.

Interestingly, the Regge symmetries not only restrict the  range
of the discrete manifold, but also allow an assignment of the
involved modes.

As a continuation of our recent work, this paper has demonstrated
that interesting insight into properties of the basic building
blocks of spin networks, the Wigner $6j$ symbols or Racah
coefficients, is gained by exploiting their self dual properties
and studying them as a function of two variables, an approach most
natural in view of their origin as matrix elements. Features of
their imaging of the orthonormal matrices are fostered by
computational advances, which is being developed on traditional
and new recurrence relations, which also allow interpretation of
the underlying Hamiltonian mechanics: the borderline of  the two
limiting modes --quasi--classical and deeply quantum-– has been
the object of this paper. The characteristic features of this
boundary line, described by the caustic curves which follow the
turning points of the classical motion, have been elucidated. A
key role was also revealed of the surprising Regge symmetries.

Further work based on these tools involves detailed studies of
some of the important aspects considered too briefly or not at all
in this paper, such as the classical, mirror and Regge symmetries,
the limiting cases to the simpler $3j$ coefficients and rotation
matrix elements, the extensions to higher $3nj$ symbols and to the
so called $q$ deformations.

Semiclassical and asymptotic analysis provide limiting
relationships converging into the Askey scheme. Relationships also
arose then with orthogonal polynomials, and indicate avenues to
views to generalizations (continuous extensions, relationships
with harmonics of rotation groups, and finally q-extensions,
\ldots). All these extensions may need some modifications when
detailed properties are discussed, but in general studying $6j$
symbols continues to deserve further work.

Use is finally pointed out for discretization algorithms of
applied quantum mechanics, particular attention being devoted to
problems encountered in atomic and molecular physics.

 \section*{Acknowledgments}
MR and ACPB acknowledge the CNPq agency for the financial support.
MR is also grateful for the financial support by the FAPESB
agency.

\vfill
\newpage

\end{document}